# Structural study of analogues of Titan's haze by trapped ion mobility coupled with a Fourier transform ion cyclotron mass spectrometer

Christopher P. Rüger[1], Julien Maillard[1,2], Johann Le Maître[1,3], Mark Ridgeway[4], Christopher J. Thompson[4], Isabelle Schmitz-Afonso[1], Thomas Gautier[2], Nathalie Carrasco[2], Melvin A. Park[4], Pierre Giusti[3], Carlos Afonso[1]

1 - Normandie Université, INSA Rouen, UNIROUEN, CNRS, COBRA, 76000, Rouen, France

2 - LATMOS/IPSL, UVSQ Université Paris-Saclay, UPMC Univ. Paris 06, CNRS, Guyancourt, France

3 - TOTAL Refining and Chemicals, TRTG Gonfreville l'Orcher, Rogerville, France

4 - Bruker Daltonics, Billerica, MA 01821, United States of America

*christopher.rueger@uni-rostock.de*

**ABSTRACT**

The aerosols present in the atmosphere of the Saturn's moon Titan are of particular planetary science interest and several spacecraft missions already allowed to gather spectroscopic data. Titan haze's analogs, so-called tholins, were produced on earth to push forward the comprehension of their formation and properties. In this study, this highly complex mixture was analyzed here for the first time by trapped ion mobility spectrometry coupled to ultra-high resolution mass spectrometry (FTICR MS). Electrospray ionization revealed the characteristic $CHN_x$-class components, with $CHN_{5-6}$ and DBE 6-7 most abundant. Deploying specialized visualization, enabled by accurate mass measurements and elemental composition assignments, the adapted Kendrick mass defect analysis highlights the $C_2H_3N$ homolog series, whereas the nitrogen-modified van Krevelen diagram exhibits a clear trend towards H/C 1.5 and N/C 0.5. More interestingly, the representation of *m/z* versus collision cross section (CCS) allowed hypothesizing a ramified N-PAH structural motif. State-of-the-art IMS is currently not able to resolve the isomeric continuum of ultra-complex mixtures; thus peak parameter other than the CCS-value are explored. As such, analyzing the mobility peak width versus *m/z* shows a linear increase in isomeric diversity between *m/z* 170 and 350 and a near plateau in diversity at higher *m/z* for the N-PAH-like structure. Due to the high complexity of the sample, these structural insights are only to be revealed by TIMS-FTICR MS.

**Introduction.** Titan, the second largest moon in our solar system, possesses a thick atmosphere mainly composed of nitrogen mixed with a few percent methane. In this gaseous environment, nitrogen radicals are created by, for example, UV radiation from the sun. Their reaction with methane leads to the formation of a thick haze. [1] Because of this particular formation, the study of this fog represents a substantial interest for prebiotic chemistry and planetary sciences. Space missions (Voyager and Cassini-Huygens) provided the first information on the chemical composition of this haze. They revealed the presence of mainly $NH_3$ and HCN chemical motif. However, due to technical limitations of the on-boarded instruments, the exact composition of this dust remains mostly unknown. [2] These limitations can be overcome, and further knowledge on the Titan atmosphere can be gained by producing analogs of the haze on earth, so-called tholins. [3]

Tholins were investigated by multiple mass spectrometric approaches, such as in coupling gas/liquid chromatography [4, 5] and various, direct infusion experiments. [6–9] Very recently, ion-mobility spectrometry coupled with a time-of-flight mass spectrometer (IMS-TOF MS) was performed to obtain structural and isomeric information. [10] By comparison with common structural elements, several structural motifs could be excluded, and an overview of potential shapes was allowed. However, the conclusions of this and other previous investigations revealed limitations due to the restricted resolution of the mass spectrometer and the extreme complexity of tholins samples. This previous investigation was limited to a relatively narrow mass range focusing on small constituents. Commonly, far heavier species are detected in tholins; thus, a significant proportion of the structural information might be lost.

The present work was motivated to improve on the prior limited view on the molecular structure of tholins by using higher resolution mass and mobility analyzers - namely trapped ion mobility spectrometry (TIMS), to Fourier transform ion cyclotron resonance mass spectrometry (FTICR MS). FTICR MS with its unbeaten mass accuracy and resolving power serves as the ideal analyzer for the ultra-complex tholins samples. TIMS-FTICR MS was recently deployed for the description of petroleum and other complex natural mixtures. [11–13] To the best of the author's knowledge, this is the first proof of principle study addressing a planetary science subject. As an example, the soluble fraction of a tholins sample is investigated by electrospray ionization (ESI) coupled to TIMS-FTICR MS. The unique information derived by ion mobility separation – *i.e.*, such as mobility peak position, width, and shape, provides more detailed structural insight than mass spectrometry alone. [13, 14]

**Experimental**

Tholins were produced according to the PAMPRE process. [3] Briefly, after a few hours of cold plasma discharge of a nitrogen/methane mixture, a brown powder, the tholins sample material, was recovered. For electrospray analysis, the sample was dissolved in methanol, stirred and filtered, extracting the soluble fraction. Directly prior analysis, a water/methanol (50/50 v-%) mixture was prepared. A mixture containing phosphazene derivatives (Agilent Tuning Mix) was used for collision-cross section calibration.

Mobility-mass analysis of the tholins soluble fraction was performed by a TIMS 12 T FTICR MS. This prototype instrument was equipped with an electrospray ionization source, and spectra were recorded in positive ion mode (5 kV, sample flow rate 120 µL/h). The nature of tholins referred to be as HCN-polymer derivative with a certain degree of aromaticity (see S11 for comparison to previous laser desorption ionization data) allow ionizing the major proportion of the tholin soluble fraction by positive mode ESI. The trapped ion mobility cell replaced the standard inlet funnel. Briefly, reduced mobility ($K_0$) was scanned from $1.21^{-4}$ to $3.37^{-4}$ $m^2s^{-1}V^{-1}$ in 72 steps. Absorption mode spectra were recorded with a dynamically harmonized ICR cell and a 0,7 s transient (full-sine apodization, one zero-fill). Consequently, a mass resolving power of over 500,000 at *m/z* 400, was achieved.

After mass calibration utilizing characteristic CHN homolog rows, each scan of the time-resolved data was peak picked (S/N 9) and exported. The exported mass spectra were processed by self-written MATLAB routines. The extracted mobilograms are fitted to an asymmetric Gauss shape, and peak parameter are exported for each individual signal. Finally, elemental compositions were assigned with the following common boundaries for tholins: $C_{1-50}H_{1-120}N_{1-15}O_{0-3}S_0$. A mass error of 1 ppm and only protonated species were considered. The mean assignment error was found to be below 100 ppb. More details on the experimental can be found in the supporting information (S1).

**Results and Discussion**

The results gave evidence of the high complexity of this artificial planetary science material, and roughly 1,300 elemental compositions could be attributed up to *m/z* 540. The complexity of tholins was previously pointed out but the previous direct infusion experiments missing the added dimension of mobility separation and, thus, the further structural information. [12] The mass spectral inset in Figure 1 emphasizes the mass spectral complexity. The ultra-high resolving power of the FTICR mass analyzer could fully resolve those close signals. Most critical for the tholins sample was a 2.6 mDa split caused by $N_6$ versus $C_4H_4O_2$ and a 4 mDa split resulted by $N_2O_2$ versus $C_5$, requiring at least a resolving power of 300k and 200k at *m/z* 400, respectively. More importantly, Figure 1 reveals the global view of the ion mobility dimension in combination with the fully resolved mass spectral pattern. Previous work performed with IMS-TOF MS was not able to resolve the increasing spectral complexity at higher *m/z*-values, and, thus, was focusing on the lower *m/z*-proportion of the tholin chemical space (Figure 1, red square). [10] The TIMS-FTICR MS coupling was able to significantly enlarge the accessible chemical space revealing CCS-values ranging from 70-170 Å² with an intensity-weighted CCS-value of 120 Å² almost in the center of the detected distribution.In an initial attempt to gain insight into tholins structural motif, the global trend of CCS versus *m/z* can be compared to CCS-values of different characteristic homolog series.

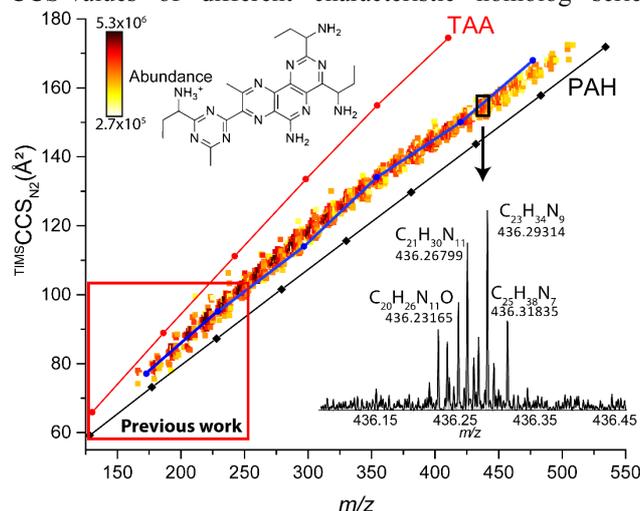

**Figure 1.** Apex collision-cross-section ($^{TIMS}CCS_{N2}$, at room temperature) versus *m/z* color-coded by abundance. The behavior of the theoretical cross-sections of different structural motif is provided for tetraalkylammonium salt species (TAA, red) and polycyclic aromatic hydrocarbons (PAH - black). A proposed structural motive for the tholins homolog series is presented in blue (archipelago ramified N-PAH). The mass spectrometric excerpt of the nominal mass 436 visualizing the high isobaric complexity. Detailed information on the homolog rows can be found in Table S1 and Figure S1.

The global trend of CCS versus *m/z* can be compared to CCS-values of different characteristic homolog series. This approach is visualized in Figure 1 for the tetraalkylammonium species (TAA) and the polycyclic aromatic hydrocarbons (PAH). Noticeably, neither series agree very well with the observed pattern and moreover an overall slightly flattening of the pattern (decrease of the slope) at higher *m/z* is revealed. Based on previous molecular level studies on tholins [5, 7, 9], we hypothesize ramified archipelago-type nitrogen-containing PAHs (N-PAHs) as the primary structural element (Figure S1). Theoretical calculations of the cross-sections for those species, given as blue line in Figure 1, revealed an excellent agreement to the experimentally observed data. This first comparison already reveals for structural information accessible only via TIMS-FTICR MS.

The elemental compositions assignments reveal $CHN_x$-class constituents are prevalent, and compositions with up to 13 nitrogen could be found (Ø#N 6.8) in this soluble methanol fraction. This observation is in agreement with previous studies, which highlight the dominant presence of pure $CHN_x$-class constituents. [4, 9] Oxidized components (artifact from the rapid oxidation when exposed to Earth atmosphere) were found to be below 15 % of the overall abundance which was an acceptable level, as the focus of this study was to proof TIMS coupled to FTICR MS as a preferable approach for the investigation of extraterrestrial relevant materials.

In contrast to the classical van Krevelen diagram visualizing H/C versus O/C, the modified representation utilizing the N/C is ideal for the study of tholins. [4] In Figure 2 the respective N/C-Van-Krevelen diagram is given for all $CHN_x$-class constituents. The pattern centered close to the respective intensity weighted values of 1.56 and 0.44 for the H/C and N/C, respectively. Species with a nitrogen-number up to equaling the carbon number were found. Noticeably, the peak apex CCS-value as a measure of the mean size of the specific molecular formula is low at the edges of the pattern and increasing towards the central point of the pattern. This behavior is caused by the stepwise addition of building blocks with an average composition of $C_2H_3N$ with N/C and H/C similar to the central point. Thus, the start of an isomeric tholin series was found to be rather non-characteristic covering a broad region space in the van Krevelen diagram, whereas with increasing *m/z* caused by the addition of $C_2H_3N$-building blocks to the molecular formula results in the size increases and consequently the CCS-value rises towards H/C 1.5 and N/C 0.5. As $C_2H_3N$ was identified as an essential building block, the respective Kendrick mass defect diagram can be drawn with the $C_2H_3N$-homolog rows at the horizontal lines (see Figure S2). The correlation pointed out in Figure 1 between *m/z* and CCS can be seen. Moreover, a less prevailing effect of increasing CCS with increasing KMD-$C_2H_3N$ is observed. Thus, different homolog rows might exhibit the same nominal mass but different size (reflected by CCS), which correlates with KMD-$C_2H_3N$. Nonetheless, this effect is weak, and we can conclude, that the size-span of the homolog rows is rather low compared to the increase in CCS with raising *m/z*.

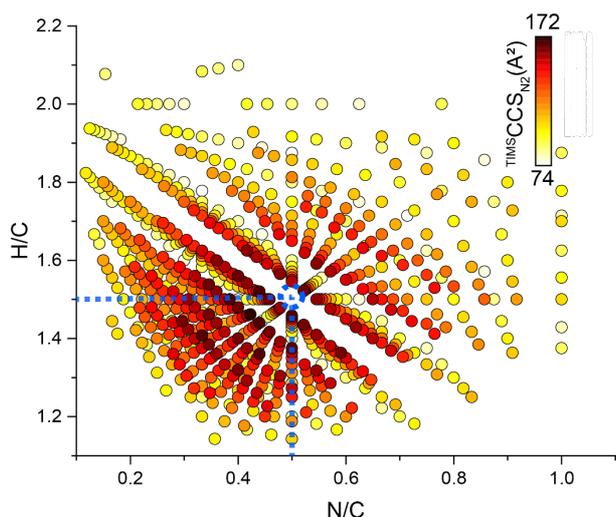

**Figure 2.** H/C versus N/C van Krevelen representation of all $C_xH_yN_z$-class constituents color-coded according to peak apex $^{TIMS}CCS_{N2}$-value. A clear center of the pattern can be seen at H/C 1.5 and N/C 0.5 (marked in blue).

Aside from the CCS, allowing to hypothesize structural aspects, the full-width-at-half-maximum (FWHM) serves as a measure for the isomeric diversity. [15] Due to time-of-flight mass spectrometric instrumentation, this approach was previously limited to relatively simple mixtures or low $m/z$ ions. TIMS-FTICR MS with significantly higher mass resolving power and accuracy allows to push these limitations drastically, and Figure 3 visualizes the mobility FWHM versus $m/z$ of the tholins signals. It can be observed that the standard tune mixture FWHM (containing a specific isomer) increases linearly, whereas the tholin signals reveal an apparent flattening of the pattern. In the lower $m/z$ range from 150-350 a linear trend can be found, whereas for the heavier constituents (> $m/z$ 350) a lower increase or eventually at above $m/z$ 400 a near plateau effect is observed. Thus, the isomeric diversity is increasing linearly for the low $m/z$ ions ($m/z$ 150-300). We hypothesize that structural changes due to increased flexibility cause the observed plateau effect of the $^{TIMS}$FWHM at higher m/z-values (starting ~ $m/z$ 370). We presume a folding effect, which was previously reported in multiple IMS-MS studies for polymers. [16–18] As tholins are referred to be HCN-polymers with carbon and hydrogen insertion, we could vaguely assume folding of larger tholin species can occur and will lower the isomeric diversity in terms of most compact and largest structure.

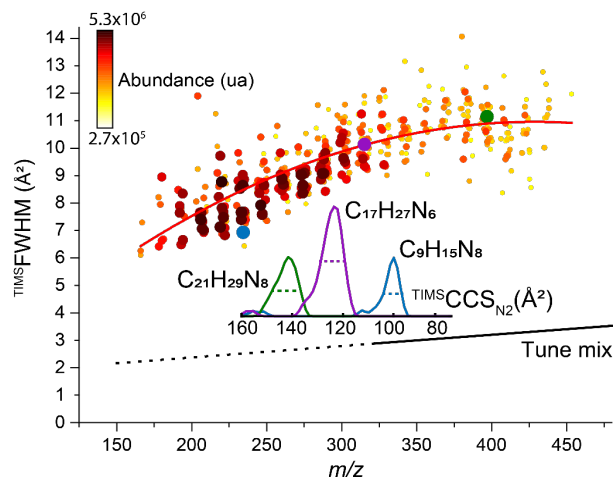

**Figure 3.** Full width at half maximum (FWHM) versus $m/z$ color- and size-coded to the summed abundance of the extracted mobilograms. Interestingly, the pattern of the FWHM is flattening towards higher $m/z$ allowing for implication of the isomeric complexity growth. As comparison, the standard mixture values are displayed and interpolated for lower $m/z$-values (dotted black line). The insert gives the mobilogram for three selected components with a low (blue), a medium (purple) and a large $m/z$ (green).

**Conclusion**

The presented study gives a proof of principle for deploying trapped ion mobility spectrometry coupled to Fourier transform ion cyclotron resonance mass spectrometry for analysis in the field of planetary science. Tholins, a highly complex artificial mixture mimicking the products on Titan, could be successfully described on the molecular level by mass spectrometry and on the isomeric level by ion mobility. Thus, IMS serves as a beneficial and complementary separation technique for FTICR MS, in particular for such complex mixtures. We hypothesized structural motif for tholins in the upper $m/z$-region. Peak shape parameters from the mobilograms are mined to gather additional information, such as the peak width as a measure for isomeric complexity. Future studies will focus on a comparison of different tholins materials, *e.g.*, solubility fractions. The utilization of the ion mobility peak shape for complex mixtures will be further investigated as a promising tool for the structural assessment. In particular, complex mixtures which are far from being resolved by IMS can be approached.

**Acknowledgment**

Thanks to the EU for funding via the ERC PRIMCHEM project (No. 636829). This work was supported at Chimie Organique Bioorganique Réactivité Analyse (COBRA) laboratory by the European Regional Development Fund (ERDF) N°31708, the Région Normandie, and the Laboratoire d'Excellence (LabEx) Synthèse Organique (SynOrg) (ANR-11-LABX-0029).